\documentclass[11pt,a4paper,english,superscriptaddress,aps,prd,preprint]{revtex4}
\usepackage{graphicx}
\makeatletter
\usepackage[latin1]{inputenc}
\usepackage{hyperref}

\newcommand{\g}{\gamma}

\newcommand{\bb}{\bibitem}
\def\pls{\partial\!\!\!/}
\def\bb{\bibitem}

\def\bs{b\!\!\!/}

\def\g{\gamma}

\def\bb{\bibitem}

\newcommand{\be}{\begin{equation}}
\newcommand{\ee}{\end{equation}}
\newcommand{\bea}{\begin{eqnarray}}
\newcommand{\eea}{\end{eqnarray}}

\begin{document}

\title{On the induction of the four-dimensional Lorentz-breaking non-Abelian Chern-Simons action}
\author{M. Gomes}
\affiliation{Instituto de F\'{\i}sica, Universidade de S\~ao Paulo\\
Caixa Postal 66318, 05315-970, S\~ao Paulo, SP, Brazil}
\email{mgomes,jroberto,ajsilva@fma.if.usp.br}
\author{J. R. Nascimento} 
\affiliation{Instituto de F\'{\i}sica, Universidade de S\~ao Paulo\\
Caixa Postal 66318, 05315-970, S\~ao Paulo, SP, Brazil}
\affiliation{Departamento de F\'{\i}sica, Universidade Federal da Para\'{\i}ba\\
 Caixa Postal 5008, 58051-970, Jo\~ao Pessoa, Para\'{\i}ba, Brazil}
\email{jroberto,passos,petrov@fisica.ufpb.br}
\author{E. Passos}
\author{A. Yu. Petrov}
\affiliation{Departamento de F\'{\i}sica, Universidade Federal da Para\'{\i}ba\\
 Caixa Postal 5008, 58051-970, Jo\~ao Pessoa, Para\'{\i}ba, Brazil}
\email{jroberto,passos,petrov@fisica.ufpb.br}
\author{A. J. da Silva}
\affiliation{Instituto de F\'{\i}sica, Universidade de S\~ao Paulo\\
Caixa Postal 66318, 05315-970, S\~ao Paulo, SP, Brazil}
\email{mgomes,ajsilva@fma.if.usp.br}

\begin{abstract}
A four-dimensional Lorentz-breaking non-Abelian Chern-Simons like action is generated as a one-loop perturbative correction via an appropriate Lorentz-breaking coupling of the non-Abelian gauge field to a spinor field. This term is shown to be regularization dependent but nevertheless it can be found unambiguously in different regularization schemes at zero and at finite temperature. 
\end{abstract}

\maketitle
During the last years, different aspects of the Lorentz symmetry breaking have 
been intensively studied \cite{Kost1}. One of the theoretical consequences of this effect is the birefringence of light in vacuum.
After the formulation of the concept of noncommutativity of the space-time \cite{SW}, 
which implies in Lorentz symmetry breaking (see the discussion in 
\cite{CarKos}), The interest in this subject has greatly increased . 
One of the implications of the Lorentz symmetry breaking is the 
possibility of introducing a lot of new  couplings in the 
Standard Model \cite{Kostel}.  These terms may arise from radiative corrections to some
Lorentz-breaking field theories at zero \cite{Ch,Ja,pv,ags,cc,al} and at finite temperature \cite{Pas1,Pas2,grig,N3}. 
Alternatively, they may be induced
from the deformation of the canonical commutation relation through the use of the noncommutative fields method \cite{Gamboa, Gamboa2}.

Recently, the renormalizability of the Yang Mills (YM) theory with a 
four-dimensional non-Abelian 
Lorentz-breaking Chern-Simons (CS) term was studied in \cite{DC}. 
The induction of such Lorentz-breaking CS term starting from a 
pure YM was  investigated 
within the noncommutative fields method in \cite{Gamboa2}.
In the present work we show how the same CS term can be induced 
through radiative corrections starting from a YM theory coupled with 
fermions in the presence of an interaction of the fermions with a constant
external field at zero and at finite temperture. 

We start with the following model which represents a 
non-Abelian generalization of the spinor electrodynamics with 
the Lorentz-breaking coupling 
\be
{\cal L}_{f}={\overline\psi}^i\left[(i \pls- m -  \gamma_{5}\bs)\delta^{ij}- g \g^{\mu}A^{a}_{\mu}(\Omega^{a})^{ij} \right]\psi^j.
\ee
Here $b_{\rho}$ is a constant four-vector.
The $A_{\mu}=A_{\mu}^a\Omega^a$ is a Yang-Mills field coupled to spinors $\psi$ which carry the isotopic indices, $\psi=(\psi^i)$, with $i$ taking values from 1 to $N$ with $N$ being the dimension of the chosen representation of the Lie algebra. The $\Omega^a=(\Omega^a)^{ij}$ are the Lie group generators in this representation satisfying the relations: $[\Omega^{a},\,\Omega^{b}]=if^{abc}\Omega^{c}$ and
${\rm Tr}[\Omega^{a}\Omega^{b}]=\delta^{ab}$.

The one-loop effective action   of the gauge field $A^a_{\mu}$ is $S_{YM}+S_{f}[b,A]$ where $S_{YM}$ is the Yang-Mills action and $S_{f}[b,A]$ can be expressed in the form of the following functional trace:
\be
S_{f}[b,A]=-i\,{\rm Tr}\,\ln(i\partial\!\!\! /- m - \gamma_5 \bs-g \g^{\mu}A^{a}_{\mu}\Omega^{a} ).
\ee
This functional trace can be rewritten as $S_{f}[b,A]=S_{f}[b]+S_{f}^{\,\prime}[b,A]$, with the first term being $S_{f}[b]=-i\,{\rm Tr}\ln(i\partial\!\!\! /- m - \gamma_5 \bs)$.
The  nontrivial dynamics is concentrated in the second term   $S_{f}^{\,\prime}[b,A]$, which is given by the  power series:
\be\label{ea}
S_{f}^{\,\prime}[b,A]=i\,{\rm Tr} \sum_{n=1}^{\infty}\frac1n
\Biggl[\frac1{i\partial\!\!\! /- m - \gamma_5 \bs}\,g \g^{\mu}A^{a}_{\mu}\Omega^{a}\Biggr]^n.
\ee
To make explicit the non-Abelian Chern-Simons term we should expand this expression up to the third order in the gauge field:
\be
\label{Ef1}
S_{f}^{\,\prime}[b,A]=S_{f}^{(2)}[b,A]+S_{f}^{(3)}[b,A]+\ldots
\ee
where
\be
S_{f}^{(2)}[b,A]=\frac{ig^{2}}{2}{\rm Tr}\bigl[\frac{1}{ i\partial\!\!\! /- m - \gamma_5 \bs}\;\g^{\mu}A^{a}_{\mu}\Omega^{a}\;\frac{1}{ i\partial\!\!\! /- m - \gamma_5 \bs}\;\g^{\nu}A^{b}_{\nu}\Omega^{b}\bigl],
\ee

\be
S_{f}^{(3)}[b,A]=\frac{ig^{3}}{2}{\rm Tr}[\frac{1}{ i\partial\!\!\! /- m - \gamma_5 \bs}\;\g^{\mu}A^{a}_{\mu}\Omega^{a}\;\frac{1}{ i\partial\!\!\! /- m - \gamma_5 \bs}\;\g^{\lambda}A^{b}_{\lambda}\Omega^{b}\;\frac{1}{ i\partial\!\!\! /- m - \gamma_5 \bs}\;\g^{\nu}A^{c}_{\nu}\Omega^{c}].
\ee
and the elipsis  stands for higher order terms in the gauge field.

Using the above expressions, it is easy now to verify that the one loop effective action expanded up to first order in $b_{\mu}$ may be written as
\be\label{Ef3}
S_{f}^{\,\prime}[b,A]=\int d^4x\;
k_{\rho}\epsilon^{\rho\mu\nu\lambda}(\partial_{\lambda}A^{a}_{\mu}A^{a}_{\nu}-\frac{2}{3}igA_{\mu}^{a}A_{\nu}^{b}A_{\lambda}^{c}f^{abc})
\ee
where $k_{\rho}$ is 
\be\label{I1}
k_{\rho}=2ig^{2}\int\,\frac{d^{4}p}{(2\pi)^{4}}\,\frac{b_{\rho}(p^{2}+3m^{2})-4p_{\rho}(b\cdot p)}{(p^{2}-m^{2})^{3}}.
\ee
This result exactly reproduces the structure of the non-Abelian Lorentz-breaking Chern-Simons term described in \cite{DC}. One can observe that the expressions (\ref{Ef3},\ref{I1}), after reduction to the Abelian case, coincide with the  known Abelian results \cite{AA,N3}. 
Apparently, there is a relation between the induced  Lorentz-breaking Chern-Simons term and Adler-Bell-Jackiw anomaly
as both situation are observed for the well known triangle graph. This issue has been discursed in Ref.\cite{pv}. Also, the interesting discussion of the problem of ambiguities in the Lorentz-breaking theories is presented in \cite{Ja}. By power counting, the momentum integral in expression (\ref{I1}) involves terms with logarithmic divergence so that different regularization prescriptions will produce diverse outcomes. Lorentz preserving regularizations, more precisely any regularization in which we can make: $p_\mu p_\nu \to \frac{g_{\mu\nu}}{D}p^2$, will produce finite results. By adopting the method of dimensional regularization \cite{HOOFT}, the above integral is promoted to $D$ dimensions and a straightforward calculation yields

\begin{eqnarray}
 \label{I3}
k_{\rho}&=&2ig^{2}b_{\rho}\int\,\frac{d^{D}p}{(2\pi)^{D}}\,\frac{1}{(p^{2}-m^{2})^{3}}\bigl[({1-\frac{4}{D}) }p^{2}+3m^{2}\bigl]\nonumber\\&=&\frac{4g^{2}\,(4-D)\,\Gamma{\bigl((4-D)/2\bigl)}}{\Gamma(3)(4\pi)^{D/2}}\,b_{\rho}=
\frac{g^{2}}{4\pi^{2}}\,b_{\rho},
\end{eqnarray}
which coincides with the result found in   \cite{AA} for the Abelian situation.  If, instead of dimensional regularization,  the integral in Eq. (\ref{I1}) is kept in four  dimensions the regularization enforced the replacement
\begin{equation}
 \label{I2}
k_{\rho}=6ig^{2}m^{2}b_{\rho}\int\,\frac{d^{4}p}{(2\pi)^{4}}\,\frac{1}{(p^{2}-m^{2})^{3}}=\frac{3g^{2}}{16\pi^{2}}b_{\rho},
\end{equation}
which now agrees with the Abelian result obtained in \cite{JK}.

To develop calculations in the finite temperature case, let us now assume that the system is in the state of  thermal equilibrium at a temperature $T=1/{\beta}$. In this case, we can use the Matsubara formalism for fermions, which consists in taking $p_{0}\equiv i\omega_n=(n+1/2)\frac{2\pi i}{\beta}$ and  replacing  the integration over  the zeroth component of the momentum by a discrete sum $(1/2\pi)\int dp_{0}\rightarrow\frac{i}{\beta}\sum_{n}$. Thus, the Eq. (\ref{Ef3}) can 
be written as
\be\label{Ef4a}
S_{f}^{\,\prime}[b,A]=\int d^4x\;k_{\rho}(\beta)\epsilon^{\rho\mu\nu\lambda}(\partial_{\lambda}A^{a}_{\mu}A^{a}_{\nu}-\frac{2}{3}igA_{\mu}^{a}A_{\nu}^{b}A_{\lambda}^{c}f^{abc}).
\ee
Hereafter all expressions are  in the Euclidean space (all greek indices run from 1 to 4). The vector $k_{\rho}( \beta)$ is given by
\be\label{I4}
k_{\rho}( \beta)=\frac{2g^{2}}{\beta}\,\sum^{\infty}_{n=-\infty}\int\,\frac{d^{3}\vec{p}}{(2\pi)^{3}}\,\frac{b_{\rho}(3m^{2}-p^{2})+4p_{\rho}(b\cdot p)}{(p^{2}+m^{2})^{3}}.
\ee
By extending the $\vec p$ integration to $d$ dimensions it follows that the time-like component of $k_\rho(\beta)$ is 
\be\label{I4.1}
k_{4}( \beta)=\frac{2g^{2}}{\beta}b_{4}\sum^{\infty}_{n=-\infty}\int\,\frac{d^{d}\vec{p}}{(2\pi)^{d}} \;\frac{3M^{2}_n-\vec{p}\;^{2}}{(\vec{p}\;^{2}+M^{2}_n)^{3}},
\ee
where $M^{2}_n=\omega^{2}_n+m^{2}$. Using the prescription of dimensional regularization \cite{HOOFT}, we have

\bea\label{I4.2}
k_{4}(\beta)&=&-\frac{ g^{2}}{2\beta}\frac{b_{4}}{(4\pi)^{d/2}}\,\bigl[d\Gamma(2-d/2)-6
\Gamma(3-d/2)\bigl]\,\sum^{\infty}_{n=-\infty}\,\frac{1}{(M^{2}_n)^{2-d/2}}\nonumber\\
&=&\frac{g^{2} b_4}{m^3\pi}\left(\frac{m}{2 \sqrt \pi}\right)^{d}}\bigl({a^{2}\bigl)^{\lambda-1/2}
\,(3-d)\,\Gamma\bigl(\lambda\bigl)\,\sum^{\infty}_{n=-\infty}\,\frac{1}{[(n+1/2)^2+a^{2}]^{\lambda}},
\eea
where $a=m \beta/2\pi$ and $\lambda=2-d/2$. At this point the following identity \cite{FO}:

\bea\label{fo}
\sum_n \frac{1}{[(n+b)^2 + a^2]^{\lambda}}= \frac{\sqrt{\pi}\Gamma(\lambda
- 1/2)}{\Gamma(\lambda)(a^2)^{\lambda - 1/2}}
+
4\sin(\pi\lambda)\int_{|a|}^\infty \frac{dz}{(z^2 - a^2)^{\lambda}}
Re\left(\frac{1}{\exp 2\pi(z + ib) -1}\right), 
\eea 
 valid for $1/2<\lambda<1$ can be used to get
\bea
k_{4}(\beta)&=&\frac{g^{2} b_4}{m^3\pi}
\left(\frac{m}{2\sqrt{\pi}}\right)^d\left\{2\sqrt{\pi}+\right.\nonumber\\
&&\left. +4\,(3-d)\bigl(a^{2}\bigr)^{\frac{3-d}{2}}\,\Gamma(\lambda)\sin(\pi\lambda)\int_{|a|}^\infty \frac{dz}{(z^2 - a^2)^{\lambda}}
Re\left(\frac{1}{ \exp 2\pi(z + ib) -1}\right)\right\}.
\eea
In $d=3$  this gives
\be
k_{4}(\beta)=\frac{g^{2}}{4\pi^{2}}\,b_{4},
\ee
i.e., the same result (\ref{I3}) without any dependence on the temperature, which  agrees with
the one obtained in \cite{Ebert} for the Abelian situation. If instead of (\ref{I4}) we use (\ref{I2}) as the starting point for the computation of finite temperature effects we get 
\be 
k_{4}( \beta) =b_4(\frac{3}{32\pi^2}+\frac{3}{16}F(a)), 
\ee 
where 
\be 
F(a)=\int_{|a|}^{\infty}dz(z^2-a^2)^{1/2} \frac{\tanh(\pi
z)}{\cosh^2(\pi z)}, \label{A1}
\ee 

\noindent has the following asymptotics:
$F(a\to\infty)\to0$ ($T\to0$) and $F(a\to0)\to{1}/{2\pi^2}$
($T\to\infty$), see Fig.\ref{fig2}.
\begin{center}
\begin{figure}[ht]
\centerline{\includegraphics[{angle=90,height=7.0cm,angle=180,width=8.0cm}]
{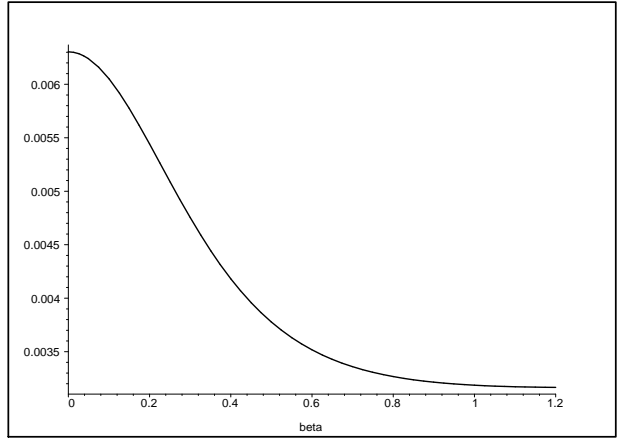}} \caption{The function $F(a)$ is different from
zero everywhere. At zero temperature ($\beta\to\infty$), the
function tends to a nonzero value $\frac{1}{2\pi^2}$.}\label{fig2}
\end{figure}
\end{center}

 Let us now consider the space part, $k_i(\beta)$, of the  vector $k_{\rho}(\beta)$. In this case, the expression (\ref{I4}) can be rewritten as:
\bea
\label{I4.1.1a}
k_i( \beta)=\frac{2g^{2}}{\beta}\sum^{\infty}_{n=-\infty}\frac{d^{3}\vec{p}}{(2\pi)^{3}}\,\frac{b_i(3m^{2}-p^{2})+4p_i(b\cdot p)}{(p^{2}+m^{2})^{3}},
\eea
 Then, considering this expression formally in $d$ space dimensions, we can replace $p_i p_j$ by $\frac{\vec{p}^2}{d}\delta_{ij}$, hence we get
\bea
\label{I4.1.1b}
k_i( \beta)=\frac{2g^{2}}{\beta}b_i\sum^{\infty}_{n=-\infty}\frac{d^d\vec{p}}{(2\pi)^d}\,\frac{4m^2-(\frac{d-4}{d})\vec{p}^2-M^2_n}{(\vec{p}^2+M^2_n)^{3}},
\eea
which now furnishes
\bea
\label{I4.1.1c}
k_i(\beta)&=&\frac{4m^2g^2}{\beta}b_i\frac{\Gamma(3-\frac{d}{2})}{(4\pi)^{d/2}}\sum^{\infty}_{n=-\infty}\,
\frac{1}{(m^2+\omega^2_n)^{3-\frac{d}{2}}}=\nonumber\\
&=&\frac{g^2 b_i}{2 m \pi^3}\left(\frac{m}2\right)^{d/2} (a^2)^{\lambda - 1/2}\Gamma\bigl(\lambda\bigl)\,\sum^{\infty}_{n=-\infty}\,\frac{1}{[(n+\frac{1}{2})^2+a^2]^{\lambda}},
\eea
where we have introduced $\lambda=3-\frac{d}{2}$. We cannot apply the relation (\ref{fo}) for $d=3$, because the
integral in that expression  does not converge. Thus, let us perform the
analytic continuation of that relation; we obtain  \cite{grig}
\bea
&&\int_{|a|}^\infty \frac{dz}{(z^2 - a^2)^{\lambda}}
Re\left(\frac{1}{\exp 2\pi(z + ib) -1}\right)= 
\frac{1}{2a^2}\frac{3-2\lambda}{1-\lambda}
\int_{|a|}^\infty \frac{dz}{(z^2 - a^2)^{\lambda-1}}\times\\&\times&
Re\left(\frac{1}{\exp 2\pi(z + ib) -1}\right)-\nonumber\\  &-&
\frac{1}{4a^2}\frac{1}{(2-\lambda)(1-\lambda)} \int_{|a|}^\infty
\frac{dz}{(z^2 - a^2)^{\lambda-2}}
\frac{d^2}{dz^2}Re\left(\frac{1}{\exp 2\pi(z + ib) -1}\right).\nonumber 
\eea
Thus for $d=3$ the Eq. (\ref{I4.1.1c}) takes the form 
\be 
k_{i}( \beta) =b_i(\frac{1}{4\pi^2}+\frac{1}{2}F(a)), 
\ee 
where $F(a)$ was defined in (\ref{A1}).
 Thus, we see that at high
temperature the Chern-Simons coefficient is twice its value at zero
temperature, i.e., $k_{i}(\beta\to0)=\frac{1}{2\pi^2}$. 
On the other hand, at zero temperature, one recovers the result $k_{i}(\beta\to\infty)=\frac{1}{4\pi^2}$.


We have generated the non-Abelian Lorentz-breaking Chern-Simons term via the Lorentz-breaking coupling of the Yang-Mills field with the spinor 
field at zero and at finite temperature. The essential property of the result is that within the framework of  dimensional regularization this term turns out to be finite. We note that the derivative expansion approach naturally allows to preserve the gauge invariance for the quantum corrections.  It is natural to expect that at least some of other Lorentz-breaking terms which existence was predicted in \cite{Kostel} also can be generated via appropriate couplings of the gauge or gravity fields with some matter fields. 
 
We have also obtained the coefficient $k_\rho$  for the non-Abelian Lorentz-breaking Chern-Simons term at the finite temperature. We found that the results for this term turn out to be dependent on the regularization scheme both at zero and at finite temperature   (in a particular regularization scheme the time-like component was found to be temperature independent). Considering the dependence on the regularization scheme, one should note that the momentum integral  determining the value of the vector $k_{\rho}$ is formally superficially divergent, thus dependence of its finite part on the renormalization procedure is very natural. However, in the regularization schemes suggested in the paper the divergent part identically disappears as a consequence of the rotational invariance of the relevant integrands.

{\bf Acknowledgements.} This work was partially supported by Conselho Nacional de Desenvolvimento Cient\'{\i}fico e Tecnol\'{o}gico (CNPq) and Funda\c c\~ao de Amparo \`a Pesquisa do Estado de S\~ao Paulo (FAPESP). The work by A. Yu. P. has been supported by CNPq-FAPESQ DCR program, CNPq project No. 350400/2005-9.

\end{document}